\documentclass[journal,10pt,twocolumn]{IEEEtran}
\usepackage{epsfig}
\usepackage{graphicx}
\usepackage{subfigure}
\usepackage{amssymb}
\usepackage{makeidx}
\usepackage{amsmath}
\usepackage{amsthm}
\usepackage{graphicx}
\usepackage{epsf}
\usepackage{epsfig}
\usepackage{psfig}
\usepackage{ccaption}
\usepackage{array}
\usepackage{tabularx}
\usepackage{multirow}
\usepackage{epsfig}
\usepackage{cite}
\usepackage{procedure,algorithm,algorithmic}

\newtheorem{proposition}{Proposition}
\newtheorem{lemma}{Lemma}

\newtheorem{remark}{Remark}

\begin{document}

\title{Joint Communications and Sensing: \\A Comprehensive Study}
\author{Husheng Li
\thanks{H. Li is with the Department of Electrical Engineering and Computer Science, the University of Tennessee, Knoxville, TN (email: husheng@eecs.utk.edu, phone number: 865-974-3861, fax number: 865-974-5483,  address: 310 Ferris Hall, 1508 Middle Drive, Knoxville, TN, 37996). }}
\maketitle

\begin{abstract}
Joint communications and sensing (JCS) can improve the efficiency of power, bandwidth and hardware usage. The conflict between communications and sensing is analyzed in terms of bandwidth and power. It is found that the bandwidth is approximately partitioned between communication and sensing, thus making an almost zero-sum game, while the conflict in the power is marginal. The same conclusion holds when the receiver and target are separated. The feasible region and tradeoff boundary of JCS are obtained, based on which the adaptive signaling in JCS is studied. 
\end{abstract}

\section{Introduction} 
In recent years, more focus has been paid to the joint communications and sensing (JCS), namely the integration of traditional communication and radar systems \cite{HanL2013,Paul2017}. A major motivation is the development of various cyber physical systems (CPSs), such as autonomous driving and unmanned vehicular (UAV) networks. In such CPSs, both communications and sensing are needed for the decision of control actions that need the information of the environment. Such information can either be exchanged by the CPS nodes using communications, such as broadcasting the position and velocity information that are obtained from local measurement (e.g., using GPS), or be obtained from sensing, such as radar ranging and Doppler based velocity estimation. 

\begin{figure}
  \centering
  \includegraphics[scale=0.4]{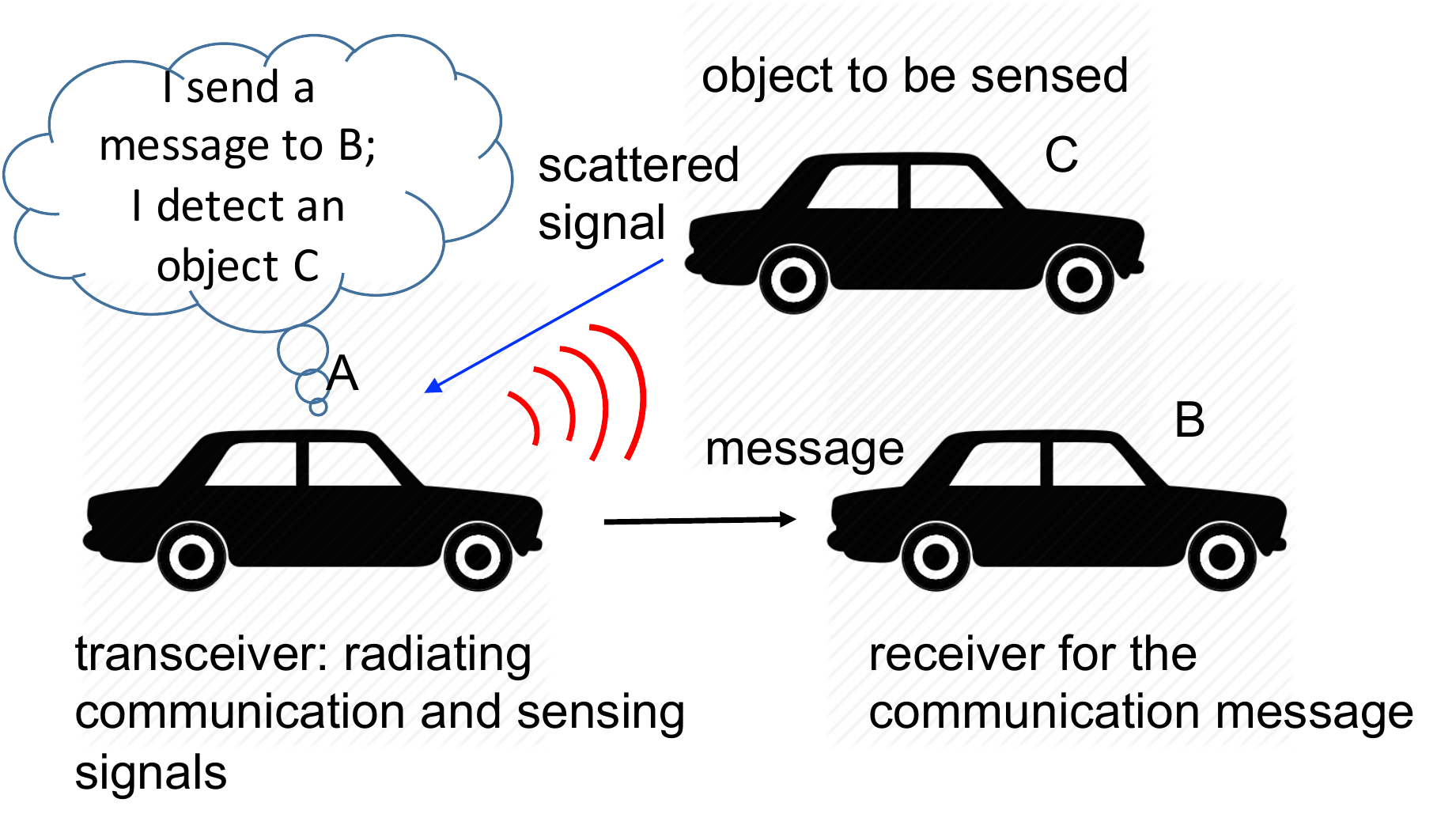}
  \caption{An example of JCS in vehicular networks}\label{fig:example}
  \vspace{-0.1in}
\end{figure}

In traditional systems, communications and radar are developed, designed and implemented almost independently, although they share some common theoretical foundation and hardware components. However, for most cases, they both use electromagnetic (EM) wave as the medium. Therefore, they can share the same waveform and complete the tasks of communications and sensing in one round of EM wave emission. An example is shown in Fig. \ref{fig:example}, in which vehicle A sends out an EM pulse, which consists of a message to vehicle B using modulations. When the message is delivered to vehicle B, the EM wave may be reflected by vehicle C (which could coincide with vehicle B). The reflected EM wave is intercepted by vehicle A and is then used to infer the information of vehicle C, which is essentially the sensing procedure. Such a JCS procedure can be understood in the following two different ways:
\begin{itemize}
\item The communication message is embedded in a sensing signal via modulation.
\item A communication signal is used to illuminate an target for sensing, similarly to a radar beam.
\end{itemize}
Moreover, the JCS can be considered as a two-way communication channel: in the forward channel, the transmitter `pushes' the information to a receiver, while in the reverse channel, it `pulls' the information from an target. 

Although there have been plenty of studies on JCS, which will be detailed in the next section, there lacks systematic study on the fundamental aspect of JCS. A key question is: \textit{how do the purposes of communications and sensing conflict with each other}? In the frequency division multiplexing (FDM) scheme of JCS, in which the communication and sensing use different frequency bands in the signal, the conflict is in the allocations of bandwidth and power. However, it is no longer obvious when the radar and communication use the same waveform. Understanding the conflict between communications and sensing is of key importance for assessing the tradeoff between the two tasks and the corresponding system design, similarly to many other tradeoff problems in communications, such as multiplexing-diversity tradeoff in multi-antenna systems \cite{Zheng2003}. 

In this paper, we study the fundamental tradeoff between communications and sensing, based on frequency modulation waveforms that are widely used in wideband radar systems. Through mathematical analysis, we conclude that (a) the signal bandwidth is approximately partitioned between communications and sensing; hence, the shared waveform scheme is similar to FDM or time division multiplexing (TDM) scheme of JCS; the same conclusion holds when the receiver and sensed targeted are separated; (b) communication and sensing have marginal conflict in the power when the receiver and sensed target are separated, thus making the shared waveform scheme more efficient than the FDM and TDM schemes. The feasible region and optimal tradeoff for JCS will be obtained, under the guidance of which we will study the adaptive signaling in JCS. 

The remainder of this paper is organized as follows. The existing studies related to this paper will be detailed in Section \ref{sec:related}. The signal model, as well as the modulation scheme, will be discussed in Section \ref{sec:model}. Then, the performance metrics of JCS will be analyzed in Section \ref{sec:performance}, based on which the tradeoff of communications and sensing will be obtained in Section \ref{sec:tradeoff}. An adaptive signaling scheme will be proposed in Section \ref{sec:adaptive}. The numerical results and conclusions will be given in Sections \ref{sec:numerical} and \ref{sec:conclusion}, respectively. 

\section{Related Work}\label{sec:related}
\subsection{Joint Communications and Sensing:} In this case, the same EM waveform is shared by communications and sensing, namely the signal carries communication messages to the receiver and the reflected signal carries the information of target for sensing. Comprehensive surveys can be found in \cite{HanL2013,Paul2017}. The integration of communications and radar has been studied for spread spectrum and multicarrier signals \cite{Sturm2011}, for stepped frequency waveforms \cite{YangJ2012}, for OFDM signals \cite{Sturm2011}, for WiMax signals \cite{Arroyo2013}, for frequency hopping MIMO signals \cite{Hassanien2017}, and for shift keying signals \cite{Tedesso2018}. It has been exploited for traffic safety in \cite{Quijada2011}. The designs of waveform and beam pattern have been studied in \cite{Hassanien2016} and \cite{Hassanien2016_2}, respectively. In the mmWave band, the joint design has been analyzed from the signal processing perspective in \cite{Mishra2019}. In \cite{YouHan2016}, the sharing of 79GHz band for radar and communications in vehicles is studied in terms of time and frequency duplexing. A testbed combining radar and communications in the mmWave band has been built in UT-Austin \cite{Heath2016,Kumari2018}, mainly for the purpose of automotive applications. In \cite{JZhang2018}, a joint communication and radar scheme is proposed based on duplexing in different beams. {It differs from the proposed research by its dedicated hardware and waveform.}

\subsection{Sensing Waveform for Communications}: The typical radar waveforms, particularly the FMCW (chirp) waveform, have been leveraged to convey information \cite{Reynders2016,Kellett2017,Dokhanchi2019,Dwivedi2019,Barrenechea2007}. The main approach is to consider the FMCW waveform as the carrier, and the information is modulated similarly to the quadrature amplitude modulation (QAM). The author has carried out experiments using chirp radar signals in the mmWave band for communications \cite{FanYa2019}, in which the communication rate is in the order of hundreds of kbps.

\section{Signal Model}\label{sec:model}
In this section, we introduce the signal model of JCS, using traditional radar waveforms.

\subsection{System Model}
We assume that a transmitter sends out signal $s(t)$, which is received by a communication receiver and reflected by a target as well.
For simplicity, we assume that the receiver is a portion of the target; e.g., a vehicle sends a message to another vehicle while sensing the information of the same vehicle. For the cases in which the communication receiver is in a location different from the target, the major concern is the beamforming which directs different portions of power to the receiver and target. 
 
\subsection{Carrier Signal} 
We assume that the traditional radar waveform is used as the carrier, over which the communication information is modulated. With one radar pulse, the transmitted signal is given by
\begin{eqnarray}\label{eq:carrier}
s(t)=A(t)\exp\left(2\pi \int_{-\infty}^t f(\tau)d\tau+\theta_0\right), \mbox{  }t\in [0,T_p],
\end{eqnarray}
where $T_p$ is the pulse duration, $A$, $f$ and $\theta_0$ are the amplitude, frequency and initial phase, respectively. 

The following two types of radar waveforms are considered in this paper:
\begin{itemize}
\item Frequency Modulation Continuous Waveform (FMCW): In the FMCW case, the instantaneous frequency increases linearly with time, namely
\begin{eqnarray}
f(t)=St+f_0,\qquad t\in [0,T_p],
\end{eqnarray}
where $S$ is the frequency increasing slope and $f_0$ is the initial frequency. 

\item Step-Frequency (SF): In the SF scheme, the frequency is piecewisely constant and increases after each interval of constant frequency, namely
\begin{eqnarray}
f(t)=\Delta f k+f_0,\mbox{   } t\in [(k-1)\Delta t,k\Delta t],
\end{eqnarray}
for $k=1,...,K$, where $\Delta f$ is the increase of frequency between successive intervals, and $K$ is the total number of intervals, such that $K\Delta t=T_p$.
\end{itemize}
We assume that both the FMCW and SF schemes use the same bandwidth, namely $ST_p=\Delta fK$.

\subsection{Wideband JCS Signal: Modulation Scheme} 
Based on the carrier signal model in (\ref{eq:carrier}), we consider the following generic signal model with information modulation:
\begin{eqnarray}
s(t)&=&\sum_{n=0}^{N_s-1} A_n(t-(n-1) T_s)\nonumber\\
&\times&\exp\left(2\pi f_n(t-(n-1) T_s)+\theta_n\right),
\end{eqnarray}
where $N_s$ is the total number of information symbols in one radar pulse, and $T_s$ is the symbol period given by $T_s=\frac{T_p}{N_s}$. Similarly to modern communication systems, the information can be modulated in the following parameters of sinusoidal signals:
\begin{itemize}
\item Complex Amplitude: The complex amplitude $A$ (including the signal magnitude and phase) can be used to convey information, similarly to the amplitude modulation (PAM) in digital communications. However, the variation of amplitude will decrease the average SNR of reflected sensing signal, thus impairing the performance of sensing.

\item Frequency: The frequency can also be varied for conveying information, in both the FMCW and SF schemes, similarly to frequency shift keying (FSK) in digital communications. It adds randomness to the frequency, thus may affect the performance of sensing.
\end{itemize}

In this paper, we propose two modulation schemes, based on the wideband radar signal waveforms, by following the same principle of typical digital modulations. There are totally four possible combinations of modulation (amplitude and frequency) and carrier waveform (FMCW and SF). For simplicity, we will consider only the schemes of QAM-FMCW and FSK-SF.

\subsubsection{QAM-FMCW} We consider modulating the amplitude and phase, similarly to the quadratic amplitude modulation (QAM) in traditional digital communication systems. In a contrast, with a time-varying frequency the carrier in the QAM-FMCW scheme is not a pure sinusoidal function, due to the requirement of wideband radar sensing. Therefore the transmitted signal is given by
\begin{eqnarray}
s(t)&=&\sum_{n=0}^{N_s-1} A_{n}(t-nT_s)\exp\left(j2\pi \theta(t)\right),
\end{eqnarray}
where $A_n$ is the complex symbol similar to QAM, and $\theta(t)$ is the deterministic (thus carrying no information) phase determined by the FMCW scheme, which is given by
\begin{eqnarray}
\theta(t)=\int_0^t f(s)ds=\frac{1}{2}St^2+f_0t.
\end{eqnarray}

\subsubsection{FSK-SF} 
In the FSK-SF scheme, the amplitude $A$ is a constant. For assuring the performance of communications, we assume $K\gg N_s$ and $K/N_s$ is an integer. We can allow some randomness in the frequency of each symbol period, namely the frequency at time $t$ is given by
\begin{eqnarray}
f(t)=(k-1)\Delta f+f_0+\frac{m_{\lfloor k/N_s\rfloor}\Delta f}{M},
\end{eqnarray}
for $t\in [(k-1)\Delta t,k\Delta t]$, $k=1,...,K$, where $m\in \{0,...,M-1\}$ is a random variable that carries the information, and each symbol carries $\log_2M$ bits.

\subsection{Received Signals}
We first consider the received signal at the radar transceiver for the purpose of sensing. The elapsed time between the transmission and reception of the JCS signal is given by $t_d=\frac{2d}{c}$, where $d$ is the distance between the transmitter and the target, and $c$ is the light speed. For simplicity of analysis, we assume that the target is stationary, thus waiving of the Doppler shift. Then, the received reflected signal is given by
\begin{eqnarray}\label{ref:rec_radar}
r(t)&=&\sum_{n=0}^{N_s-1} A_{n}(t-nT_s)\left(t-(n-1) T_s-\frac{2d}{c}\right)\nonumber\\
&\times&\exp\left(j2\pi \theta\left(t-\frac{2d}{c}\right)+\theta_n\right),
\end{eqnarray}

The signal received by the communication receiver has the same form as (\ref{ref:rec_radar}), except for replacing $d$ with $\frac{d}{2}$, since we assume that the communication receiver is colocated with the target. We assume that the communication receiver is perfectly time synchronized with the JCS transmitter, such that the traveling time $\frac{d}{2c}$ can canceled at the mixer. 



\section{Joint Communications and Sensing}
In this section, we introduce the algorithms for communication demodulation and sensing in the JCS scheme.

\subsection{Demodulation in Communications}
We first introduce the demodulation at communications receiver, for the QAM-FMCW and FSK-SF schemes, respectively. As will be seen, the radar carrier can be perfectly removed, provided that the time is perfectly synchronized. Then the baseband signal can be used for detecting the communication message, as in traditional communication receivers. 
 
\subsubsection{QAM-FMCW} For the QAM modulation with FMCW carrier, the demodulation is similar to that of traditional linear modulation with sinusoidal carrier. Due to the assumption of perfect time synchronization, the input signal is mixed with local FMCW signal and the output is passed through a low pass filter. The procedure is given by
\begin{eqnarray}\label{eq:QAM_output}
r(t)&=&\sum_{n=0}^{N_s-1} A_{n}(t-nT_s)\exp\left(j\theta(t)\right)\exp\left(-j\theta(t)\right)\nonumber\\
&=&\sum_{n=0}^{N_s-1} A_{n}(t-nT_s),
\end{eqnarray}
which is then put in the decision maker of traditional demodulator. Here the perfect cancellation of the carrier phase is due to the assumption of perfect time synchronization. 

\subsubsection{FSK-SF} The received signal is mixed with a local mixer with the proper center frequency $(k-1)\Delta f+f_0$, namely
\begin{eqnarray}
&&\exp\left(j2\pi\left((k-1)\Delta f+f_0+\frac{m\Delta f}{M}\right)t\right)\nonumber\\
&\times&\exp\left(-j2\pi\left((k-1)\Delta f+f_0\right)t\right)\nonumber\\
&=&\exp\left(j2\pi\frac{m\Delta f}{M}t\right),
\end{eqnarray}
where the output is based on the assumption of perfect time synchronization. Then, the frequency of the mixer output is estimated and the corresponding $m$ is obtained. Traditional demodulation approaches of FSK can be employed.

\subsection{Radar Ranging} \label{subsec:ranging}
Now, we study the ranging algorithm using the reflected radar waveform. In contrast to traditional radar ranging algorithms, the major challenge here is how to remove the impact of modulated information. 

\subsubsection{QAM-FMCW}
The received signal at the radar receiver is mixed with local FMCW signal (without modulation) and the output is given by
\begin{eqnarray}\label{eq:beat}
r(t)=\sum_{n=0}^{N_s-1} A_n\left(t-(n-1)T_s-\tau)\right)\exp\left(j2\pi\tau St\right),
\end{eqnarray}
where $\tau=\frac{2d}{c}$ is the round trip time, which is to be estimated. 
To obtain the round trip time $\tau$, we can use the following two approaches:
\begin{itemize}
\item Frequency domain estimation: Notice that $A_n$ is known to the radar receiver, thus being deterministic, since it is transmitted by the same transmitter. Hence, we can carry out Fourier transform for the mixer output in (\ref{eq:beat}) and obtain
\begin{eqnarray}\label{eq:beat_freq}
\hat{r}(jw)&=&\sum_{n=0}^{N_s-1} \hat{A}_n(j(w-2\pi\tau S))\nonumber\\
&\times&e^{-j(w-2\pi\tau S)((n-1)T_s+\tau)}\nonumber\\
&=&F(j(w-2\pi \tau S)),
\end{eqnarray}
where $F=\hat{A}_n(jw)e^{-jw((n-1)T_s+\tau)}$.

When the mixer output $\hat{r}$ is contaminated by Gaussian noise, the maximum likelihood estimation of $\tau$ is given by
\begin{eqnarray}\label{eq:FMCW_ML}
\tau_{ML}=\arg\min_{\tau}|\hat{r}(jw)-F(j(w-2\pi\tau S))|^2.
\end{eqnarray}

\item Carrier synchronization: We can also consider the signal in (\ref{eq:beat}) as the carrier with frequency $\tau S$ modulated by the information symbols $A_n$. For a typical case of $d=200m$ and $S=30MHz/us$, the beat frequency $\tau S=40MHz$. If the symbol frequency (e.g., 1MHz) is much smaller than $\tau S$, we can employ the carrier synchronization approach to estimate $\tau$. When $N_s$ is sufficiently large, we can calculate the power spectrum density, estimate the center frequency $\hat{f}_c$, and calculate $\hat{\tau}=\frac{\hat{f}_c}{S}$.
\end{itemize}

\subsubsection{FSK-SF} 
\begin{figure}
  \centering
  \includegraphics[scale=0.5]{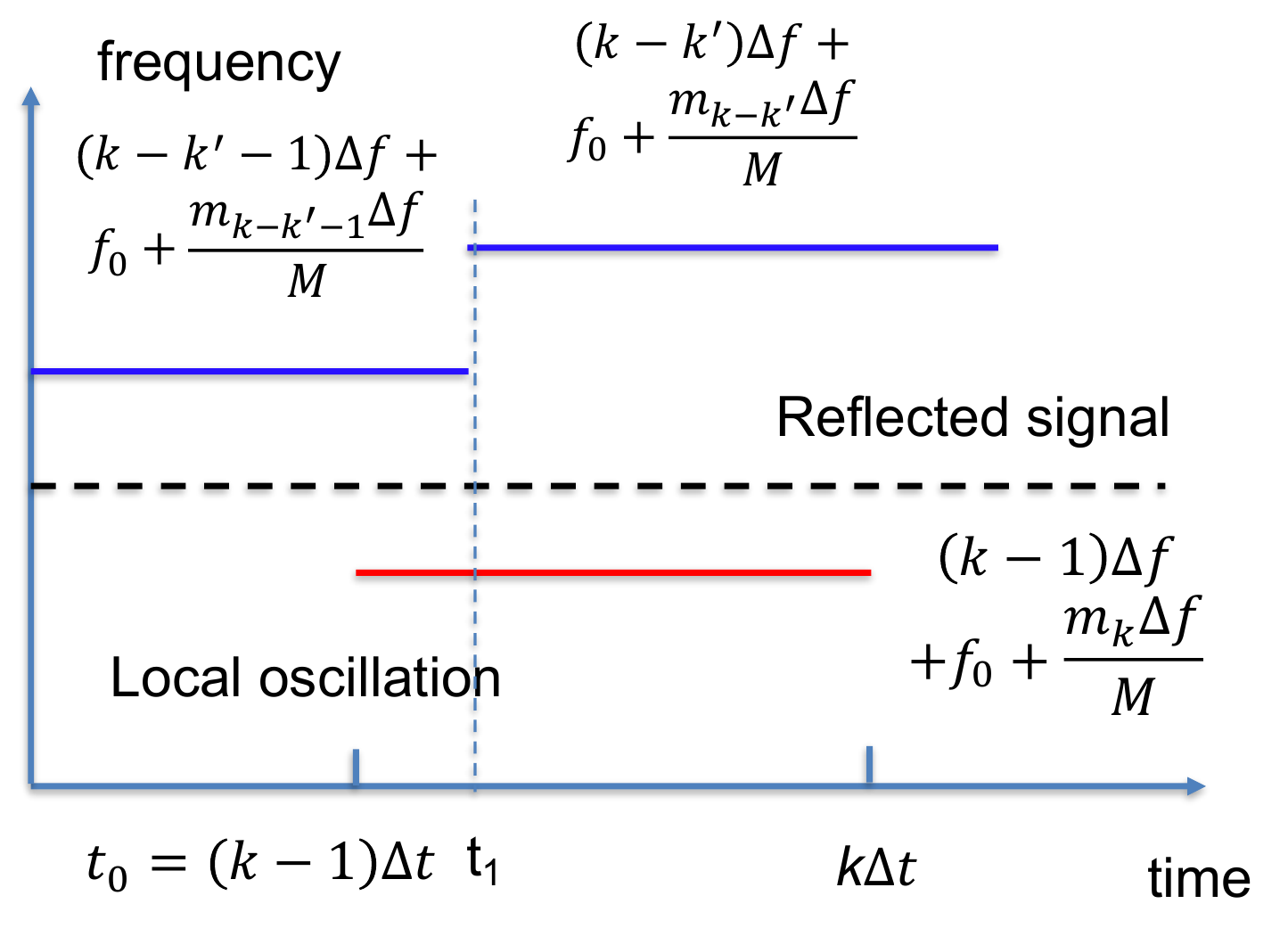}
  \caption{Timing in the mixer output of FSK-SF}\label{fig:timing}
  \vspace{-0.1in}
\end{figure}

Since the radar receiver knows the information symbol $m$ and thus the frequency, it mixes the received signal with local oscillation. The output of the mixer and filter is given by
\begin{small}
\begin{eqnarray}\label{eq:freq_out}
f(t)=\left\{
\begin{array}{ll}
&\frac{(m_k-m_{k-k'})\Delta f}{M}+k'\Delta f,\mbox{ }t\in [t_0,t_1]\\
&\frac{(m_k-m_{k-k'+1})\Delta f}{M}+(k'-1)\Delta f,\mbox{ }t\in [t_1,t_2]
\end{array}
\right..
\end{eqnarray}
\end{small}
where $t_0=(k-1)\Delta t$, $t_1=t_0+\{\frac{\tau}{\Delta t}\}$, $t_2=k\Delta t$ (where $\{\cdot\}$ means the fractional part of the number) and $k'=\left\lceil \frac{\tau}{\Delta t}\right\rceil$. The timing is illustrated in Fig. \ref{fig:timing}.

The following four steps are used to estimate $\tau$ and thus the distance:
\begin{itemize}
\item Step 1. Estimate the frequency as a function of time $t$. In the ideal case, we assume that $f(t)$ is perfectly obtained. In practice, short-time DFT can be used to estimate the frequency by selecting the peak in the spectrum.
\item Step 2. Coarse estimation: In typical setups, the term $k'\Delta f$ is significantly larger than the term $\frac{(m_k-m_{k-k'})\Delta f}{M}$, since the latter is only a fine adjustment of the frequency. Therefore, $k'$ can be estimated as $\hat{k}'=\left\lfloor\frac{f(t)}{\Delta f}\right\rfloor$.
\item Step 3. Fine estimation: Search for the optimal $t_1$ (namely the frequency change time of the received signal) such that the MSE of frequency is minimized, namely
\begin{eqnarray}
\hat{t}_1&=&\arg\min_{t_1} \int_{t_0}^{t_1}(\hat{f}(t)-f(t|\hat{k}',t_1))^2 dt\nonumber\\
&+&\int_{t_1}^{t_2}(\hat{f}(t)-f(t|\hat{k}',t_1))^2dt,
\end{eqnarray}
where $\hat{f}(t)$ is the estimated frequency using the received signal and $f(t|\hat{k}',t_1)$ is the frequency given in (\ref{eq:freq_out}) conditioned on $\hat{k}'$ and $t_1$. Note that the above integral can be replaced with summations of the frequency samples. When there are not sufficiently many samples for tracking the instantaneous frequency within one symbol period, the fine estimation of frequency cannot be accomplished.

\item Step 4. Fractional part: The estimation of the frequency change time is given by $\hat{t}_1$. Then, the fraction part of $\frac{\tau}{\Delta t}$ is given by $\hat{t}_1-(\hat{k}'-1)\Delta t$.
\end{itemize}

Then, the round trip time $\tau$ is estimated as
\begin{eqnarray}
\hat{\tau}=\hat{k}'\Delta t+\hat{t}_1-(\hat{k}'-1)\Delta t.
\end{eqnarray}

\section{Performance Analysis}\label{sec:performance}
In this section, we analyze the performance of the proposed JCS scheme. For simplicity, we focus on only QAM-FMCW. 

\subsection{Communication Performance} The exact analysis of the communication bit error rate of QAM modulation is difficult. We consider the following upper bound for the bit error rate of QAM (Eq. (4.3-30) in \cite{Proakis2007}): 
\begin{eqnarray}
P_{e,M-QAM}\leq 4Q\left(\sqrt{\frac{3\log_2 M}{M-1}\frac{P_r}{N_0}}\right),
\end{eqnarray}
where $P_r$ is the received signal and $N_0$ is the power spectral density (PSD) of the thermal noise, This upper bound becomes tight when the signal-to-noise ratio $\frac{P_r}{N_0}$ is sufficiently large. 

We leverage the following approximation:
\begin{eqnarray}
Q(x)\approx \frac{1}{12}e^{-\frac{x^2}{2}},
\end{eqnarray}
when $x$ is sufficiently large. Then, the channel capacity of the binary symmetric channel, in terms of bits per channel use is
\begin{eqnarray}
C\approx 1-\frac{e^{-\frac{x^2}{2}}}{3},
\end{eqnarray}
where $x=\sqrt{\frac{3\log_2 M}{M-1}\frac{P_tG}{N_0N_s}}$, $P_t$ is the transmit energy of the radar pulse and $G$ is the channel gain.
Therefore, the data throughput is given by
\begin{eqnarray}
T\approx\frac{N_s}{T_P}\left(1-\frac{e^{-\frac{x^2}{2}}}{3}\right)\log_2M,
\end{eqnarray}
since the channel use lasts time $T_s$. 

\subsection{Radar Performance}

The performance of FMCW radar has been analyzed in \cite{ChenWei2000,Scheer1993}. However, the SNR is assumed to be sufficiently high therein; therefore the performance is dependent on only the frequency increasing slope and the pulse duration, thus being independent of the signal power. Moreover, the impact of QAM symbols (which change the amplitude of the FMCW carrier in a random manner) is novel. In the context of JCS, it is difficult to directly analyze the performances of the proposed schemes of frequency domain estimation and carrier synchronization in Section \ref{subsec:ranging}. 

In this paper, we follow the approach in the following lemma on the frequency estimation in sinusoidal signals contaminated by noise, in order to find the lower bound of the estimation error variance.
\begin{lemma}[p.57 of \cite{Kay1993}]\label{lem:Kay}
For a sinusoidal function with frequency $f$, SNR $\gamma$ and $N$ samples, the Cramer-Rao bound is given by
\begin{eqnarray}
V[f]\geq \frac{12}{(2\pi)^2 \gamma N(N^2-1)}\approx \frac{12}{(2\pi)^2 \gamma N^3}
\end{eqnarray}
\end{lemma}

The approach is based on the Fisher Information matrix defined by
\begin{eqnarray}
I_{ij}=E\left[\left(\frac{\partial \log f(x,\mathbf{\theta})}{\partial \theta_i}\right)\left(\frac{\partial \log f(x,\mathbf{\theta})}{\partial \theta_j}\right)^*\right],
\end{eqnarray}
where $f$ is the probability density function of the sample.  

\begin{proposition}\label{prop:MSE}
When $N$ samples are taken for the output of the mixer in the QAM-FMCW scheme, the MSE of ranging is lower bounded by
\begin{eqnarray}
MSE_d\geq \frac{3c^2}{8\pi^2S^2\gamma N(N+1(2N+1))}.
\end{eqnarray}
\end{proposition}
\begin{remark}
When the number of samples is sufficiently large, we have
\begin{eqnarray}
MSE_d\geq \frac{3c^2}{16\pi^2S^2\gamma N^3}.
\end{eqnarray}
We observe that the lower bound is irrelevant to the number of information symbols $N_s$. This implies that the modulation for communications have not impact on the Cramer-Rao bound of radar sensing performance. 
\end{remark}

\begin{proof}
When noise is taken into account, the mixer output of QAM-FMCW in (\ref{eq:beat}) is given by
\begin{eqnarray}
r(t)&=&\sum_{n=0}^{N_s-1} A_n\exp\left(j2\pi\tau St\right)+n(t)\nonumber\\
&=&\sum_{n=0}^{N_s-1} A_n\exp\left(j\phi(t)\right)+n(t)
\end{eqnarray}
where $\tau=\frac{2d}{c}$ is the round trip time, $n(t)$ is the noise, and 
\begin{eqnarray}
\phi(t)=\frac{4\pi dSt}{c}.
\end{eqnarray}
$N$ samples are taken for the signal $r(t)$, and the samples are denoted by $\{y_n\}_{n=1,...,N}$. The sampling period is $\delta t=\frac{T_P}{N}=\frac{T_P}{GN_s}$.

We calculate the Fisher's information:
\begin{eqnarray}
\frac{\partial \log f(x)}{\partial d}&=&\frac{\partial}{\partial d}\sum_{n=1}^N \frac{1}{2P_N}\left|y_n-A^{(n)} \exp\left(j\phi(t_n) \right)\right|^2\nonumber\\
&=&j\sum_{n=1}^N \frac{A^{(n)}}{P_N} n(t_n)\exp\left(j\phi(t_n)\right)\frac{\partial }{\partial d}\phi(t_n)\nonumber\\
&=&\frac{j4\pi S}{cP_N}\sum_{n=1}^N A^{(n)} n\delta t\exp\left(j\phi(t_n)\right),
\end{eqnarray}
where $A^{(n)}=A_{\left\lceil\frac{n}{G}\right\rceil}$.

Therefore, we have
\begin{eqnarray}
&&E\left[\frac{\partial \log f(x)}{\partial d}\left(\frac{\partial \log f(x)}{\partial d}\right)^*\right]\nonumber\\
&=&\frac{16\pi^2S^2}{c^2P_N}\sum_{n=1}^Nn^2E\left[(A^{(n)})^2\right]\nonumber\\
&=&\frac{8\pi^2S^2\gamma N(N+1(2N+1))}{3c^2}.
\end{eqnarray}

It concludes the proof by applying the Cramer-Rao bound.
\end{proof}

\subsection{Bandwidth and Power}\label{subsec:tradeoff}
Now, we consider the consumption of bandwidth and power for the tasks of communications and radar sensing. When separate waveforms and frequency bands are allocated to communications and radar (either dynamically or stationarily), the power and bandwidth are split to the two tasks. Our goal here is to study whether they are also split to the two tasks. 

Here we consider a generic radar waveform, which is given by
\begin{eqnarray}
s(t)=\sum_{m=-\infty}^\infty \sum_{n=1}^{N_s} A_{mn} g_n(t-(m-1)T_p),
\end{eqnarray}
where $g_n$ is the carrier waveform in the $n$-th symbol period within the same pulse, namely
\begin{eqnarray}
g_n(t)=\exp\left(j2\pi\left(\frac{1}{2}St+f_0t\right)\right)W_n(t).
\end{eqnarray}
where $W_n(t)$ is the rectangular window function between $[nT_s,(n+1)T_s]$

The information sequence $\{A_{m,n}\}$ are i.i.d, which makes the signal a cyclo-stationary random process, due to the repeated carrier waveform in different radar pulses. 

Using the same argument as in \cite{Proakis2007} (Section 3.4-2), the autocorrelation function $R(t+\tau,t)$ is given by
\begin{eqnarray}
&&R(t+\tau,t)=\sum_{m=-\infty}^\infty \sum_{n=1}^{N_s}P_s\nonumber\\
&\times&g_n(t-(m-1)T_p+\tau)g_n(t-(m-1)T_p).
\end{eqnarray}
It is easy to verify 
\begin{eqnarray}
R(t+\tau,t)=R(t+\tau+T_p,t+T_p),
\end{eqnarray}
which means that the signal is cyclo-stationary. The term $P_s$ is the power of signal. We denote it by $P_s$.

Due to the cyclostationarity, the autocorrelation function can be decomposed to the Fourier series
\begin{eqnarray}
R(t+\tau,t)=\sum_{n=-\infty}^\infty R^{n}_{T_p}(\tau)e^{ \frac{-j2\pi nt}{T_p}},
\end{eqnarray}
where the Fourier series are given by
\begin{eqnarray}\label{eq:autoco}
&&R^{m}_{T_p}(\tau)\nonumber\\
&=&\frac{1}{T_p}\int_{0}^{T_p}R(t+\tau,t)e^{\frac{-j2\pi mt}{T_p}}dt\nonumber\\
&=&\frac{P_s}{T_p}\sum_{l=-\infty}^\infty\sum_{n=0}^{N_s-1} \nonumber\\
&&\int_{nT_s}^{(n+1)T_s}g_n(t-lT_p)g^*_n(t+\tau-lT_p)e^{\frac{-j2\pi mt}{T_p}}dt\nonumber\\
&=&\frac{P_s}{T_p}\sum_{n=0}^{N_s-1} \int_{nT_s}^{(n+1)T_s-\tau}g_n(t)g_n^*(t+\tau)e^{\frac{-j2\pi mt}{T_p}}dt\nonumber\\
&=&\frac{P_s}{T_p}\sum_{n=0}^{N_s-1} g_n(\tau)\ast \left(g_n(-\tau)e^{\frac{j2\pi mt}{T_p}}\right).
\end{eqnarray}

Notice that the spectrum of $g_n(t)$ is given by
\begin{eqnarray}
\widehat{g_n}(f)&=&\widehat{gW_n}(f)\nonumber\\
&=&\text{sinc}(fT_s)e^{-j2\pi nT_sf}\ast \hat{g}(f),
\end{eqnarray}
where $\hat{\cdot}$ means Fourier transform, and the term $e^{-j2\pi nT_sf}$ is due to the time shift of the window function $W_n$. 

Then, the power spectrum density of the $m$-th order is given by
\begin{small}
\begin{eqnarray}
\mathcal{S}^{m}_{T_s}(f)&=&\sum_{n=0}^{N_s-1} \widehat{g_n}(f) \widehat{g_n}\left(f-\frac{m}{T_p}\right)\nonumber\\
&=&\frac{P_s}{T_s}\sum_{n=0}^{N_s-1}\text{sinc}\left(fT_s\right)e^{-j2\pi nT_sf} \ast \hat{g}(f)\nonumber\\
&\times&\text{sinc}\left(\left(f-\frac{m}{T_p}\right)T_s\right)e^{j2\pi nT_s\left(f-\frac{m}{T_p}\right)} \ast\hat{g}^*\left(f-\frac{m}{T_p}\right)\nonumber\\
&=&\sum_{n=0}^{N_s-1}e^{-j\frac{2\pi mn}{N_s}}\frac{P_s}{T_s}\text{sinc}\left(fT_s\right) \ast \hat{g}(f)\nonumber\\
&\times&\text{sinc}\left(\left(f-\frac{m}{T_s}\right)T_s\right)\ast\hat{g}^*\left(f-\frac{m}{T_s}\right)\nonumber\\
&=&\left\{
\begin{array}{ll}
&\frac{N_sP_s}{T_s}\left|\text{sinc}(fT_s)\ast \hat{g}(f)\right|^2,m=0\\
&0,m>0
\end{array}
\right..
\end{eqnarray}
\end{small}
where in the first equality the term $f-\frac{m}{T_s}$ is due to the term $e^{\frac{j2\pi mt}{T_p}}$ in (\ref{eq:autoco}) which results in the shift in the frequency spectrum, and the last equality is due to the fact
\begin{eqnarray}
\sum_{n=0}^{N_s-1}e^{-j\frac{2\pi mn}{N_s}}
=\left\{
\begin{array}{ll}
&N_s,m=0\\
&0,m>0
\end{array}
\right..
\end{eqnarray}

In particular, when $m=0$, which means the time average power spectrum density, we have
\begin{eqnarray}\label{eq:PSD}
\mathcal{S}^0_{T_s}(f)=\frac{P_s}{T_s}\left|\text{sinc}\left(fT_s\right) \ast \hat{g}(f)\right|^2.
\end{eqnarray}

An interesting observation on (\ref{eq:PSD}) is that the PSD of the modulated radar signal is determined by the convolution of the following two terms:
\begin{itemize}
\item The sinc function, which represents the bandwidth needed by the communications, since the data rate is inversely proportional to the symbol period $T_s$ and thus approximately proportional to the bandwidth $B_c$ of the sinc function.

\item The radar signal spectrum $\hat{g}(f)$, which determines the bandwidth $B_s$ needed for radar sensing. Note that the spectrum bandwidth of FMCW signal can be found in \cite{Klauder1960}.
\end{itemize}

Due to the property of convolution, the total bandwidth needed for the JCS is given by
\begin{eqnarray}\label{eq:bandwidth}
B_t\approx B_c+B_s,
\end{eqnarray}
which means that the total bandwidth is implicitly partitioned to the bandwidths of communications and sensing. Essentially the functions of communications and radar sensing are not sharing the bandwidth. Each degree of freedom is either used for communications, or used for radar sensing. From this viewpoint of bandwidth, the JCS scheme of shared waveform is approximately the same efficient as a FDMA or TDMA separation of communications and sensing. 

In summary, the scheme of shared waveform for JCS is similar to that of FDMA and TDMA JCS in terms of bandwidth efficiency. Meanwhile, the shared waveform is more efficient in terms of power, since the FDMA or TDMA does not utilize the reflected power in the time slot / frequency band for communications.

\section{Numerical Simulations}\label{sec:numerical}

In this section, we provide the numerical simulation results to evaluate the performance of the proposed JCS schemes. 

\subsection{QAM-FMCW}

\subsubsection{Ranging Performance}
We tested both the schemes of frequency domain estimation and carrier synchronization. The default parameters are shown in Table \ref{table:para}, where most of the radar parameters follow that of TI AWR1243BOOST radar in the 77GHz band. 

\begin{table}[]
	\centering
		\caption{Configuration of simulations for QAM-FMCW}
		\label{table:para}
	\begin{tabular}{|l|l|l|l|}
		\hline
		distance  & 100m & chirp duration $T_p$ &  60us\\
		\hline
		symbol number $N_s$ & 8 & modulation   & 16QAM \\
		\hline
		freq. inc. rate $S$  & 29.98THz/s & pulse number  & 50,000   \\
		\hline
	\end{tabular}
\end{table} 

Note that the ML estimation in (\ref{eq:FMCW_ML}) provides the asymptotically performance only when the processing is analog, namely $\tau$ is continuous. However, in practice, discrete Fourier transform is needed for the discrete-time samples. Therefore, we change the minimization to ranging all the frequency indices in a discrete manner. For the carrier synchronization scheme, we simply pick the highest power peak in the frequency spectrum of mixer output, and set the corresponding frequency as the estimation $\hat{f}_c$.

\begin{figure}
  \centering
  \includegraphics[scale=0.4]{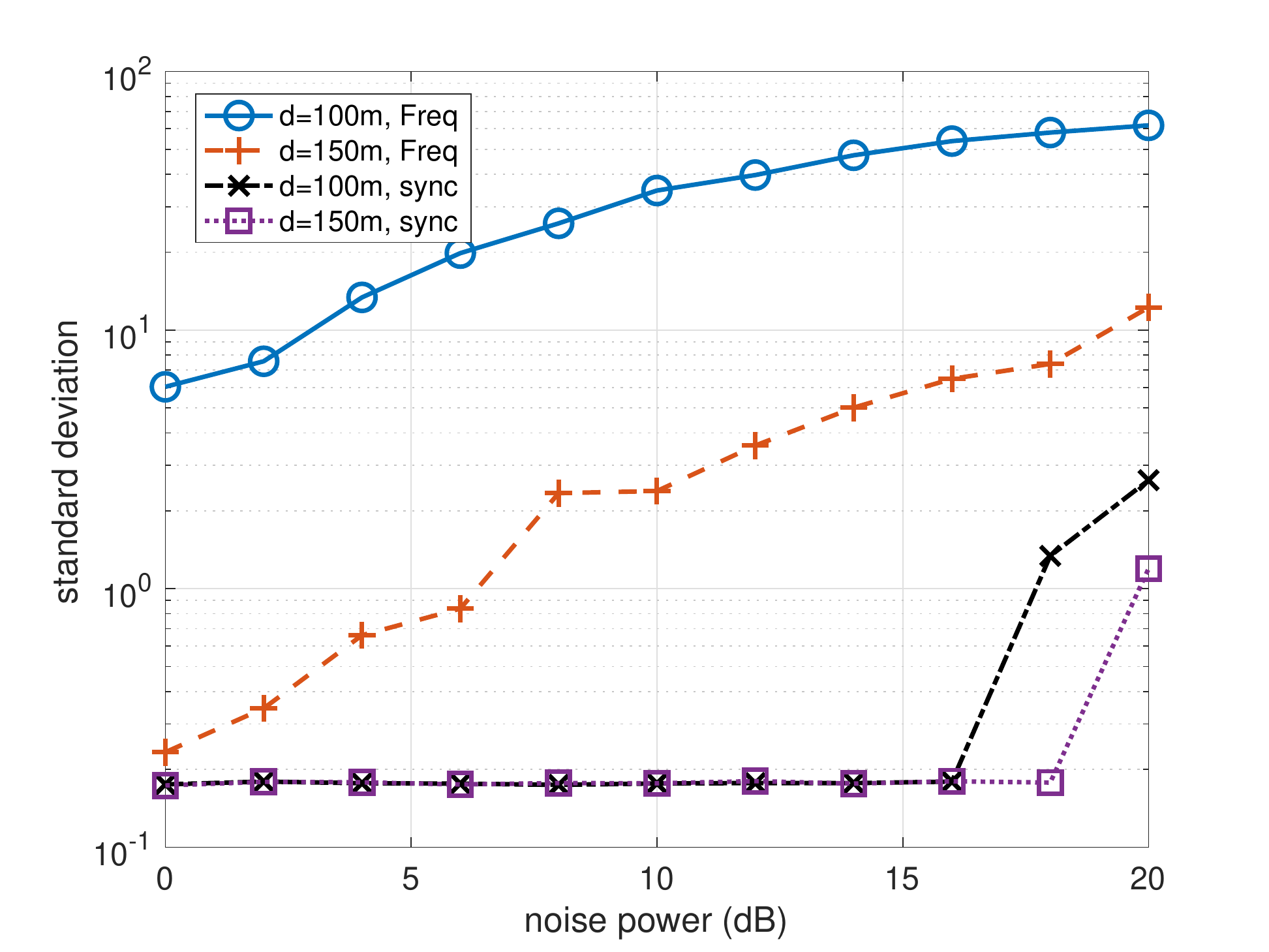}
  \caption{Ranging MSE with different noise powers and distances}\label{fig:noise}
  \vspace{-0.1in}
\end{figure}

We first tested the performance of ranging, in terms of the standard deviation of ranging errors, with respect to various noise power levels (suppose that the signal power is normalized to 1) and target distances. Both the frequency domain approach and carrier synchronization approach are tested. The results are shown in Fig. \ref{fig:noise}. An interesting observation is that the frequency domain approach performs far worse than the carrier synchronization approach. The underlying reason is that the frequency domain changes rapidly for different delays $\tau$. Since only discrete samples in the frequency domain are used, an estimation $\hat{\tau}$ close to the true value $\tau$ does not guarantee a small matching error in (\ref{eq:FMCW_ML}). In a sharp contrast, the carrier synchronization approach results in much lower ranging errors. Only when the noise power becomes much higher does the ranging error become significant. 

\begin{figure}
  \centering
  \includegraphics[scale=0.4]{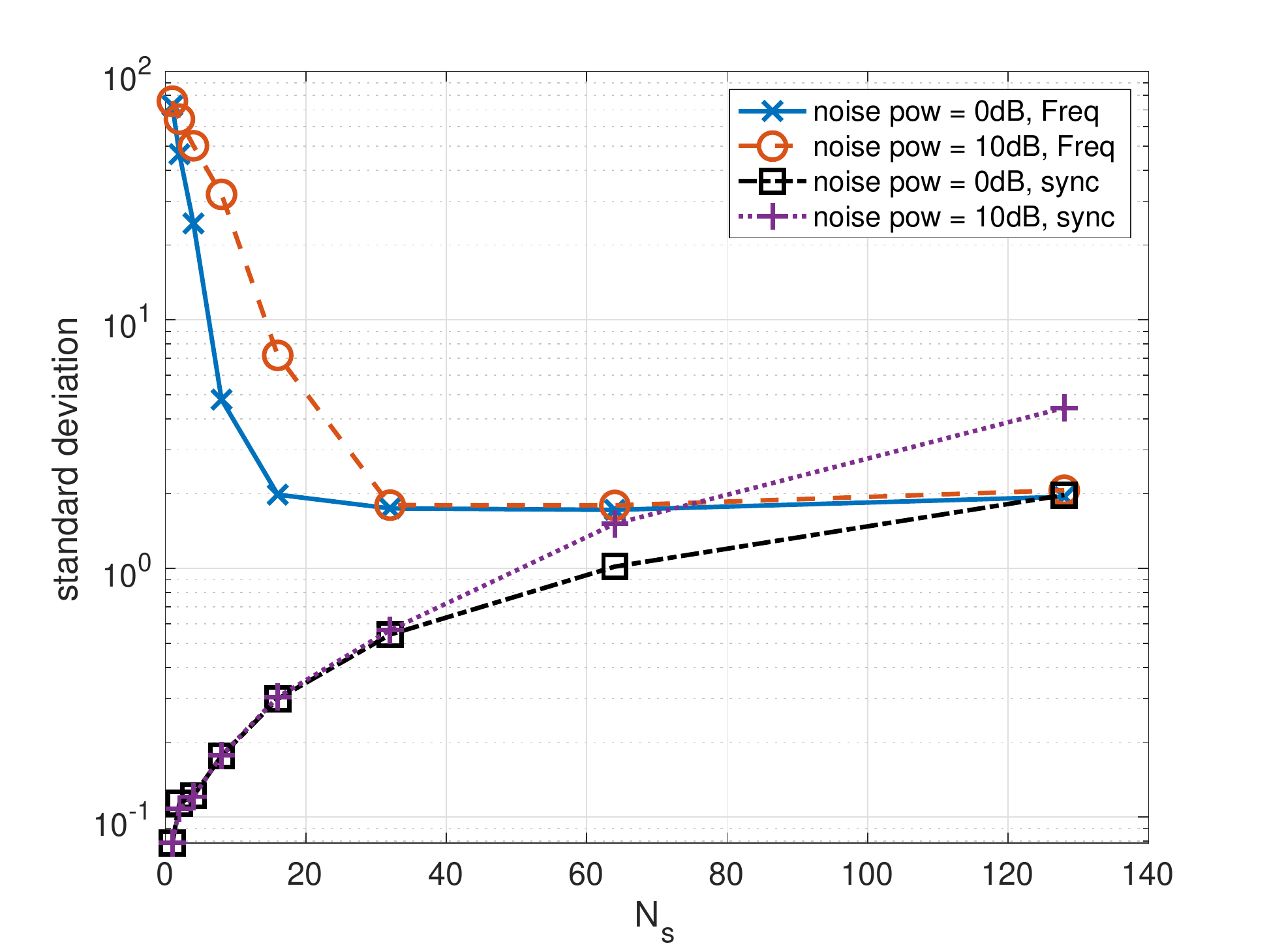}
  \caption{Ranging MSE with different noise powers and $N_s$}\label{fig:Ns}
  \vspace{-0.1in}
\end{figure}

In Fig. \ref{fig:Ns} we compare the performance of ranging with respect to different numbers of symbols in one radar pulse, namely $N_s$. We observe that, as $N_s$ increases, the performance of the frequency domain approach is improved, while that of the carrier synchronization is impaired. The detailed reason for this change is still under exploration. Since the performance of the frequency domain approach is unacceptable, we will give up this approach in the subsequent simulations. Then, the higher data rate, caused by the higher number of symbols in one chirp pulse, will result in a negative impact on the performance of ranging, thus yielding a conflict of interest between communications and radar. 

\begin{figure}
  \centering
  \includegraphics[scale=0.4]{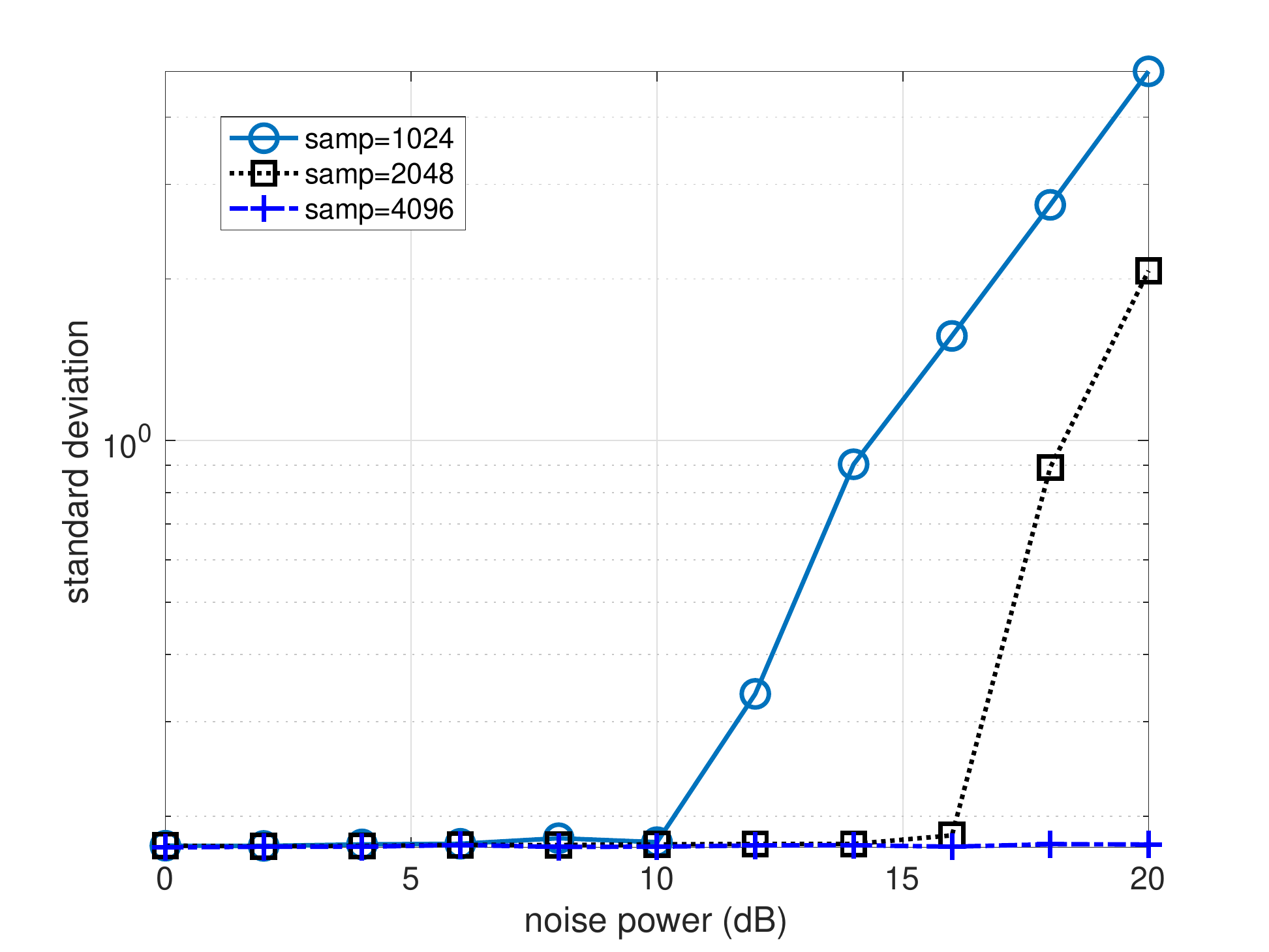}
  \caption{Ranging MSE with different noise powers and sampling frequencies}\label{fig:samp}
  \vspace{-0.1in}
\end{figure}

\begin{figure}
  \centering
  \includegraphics[scale=0.4]{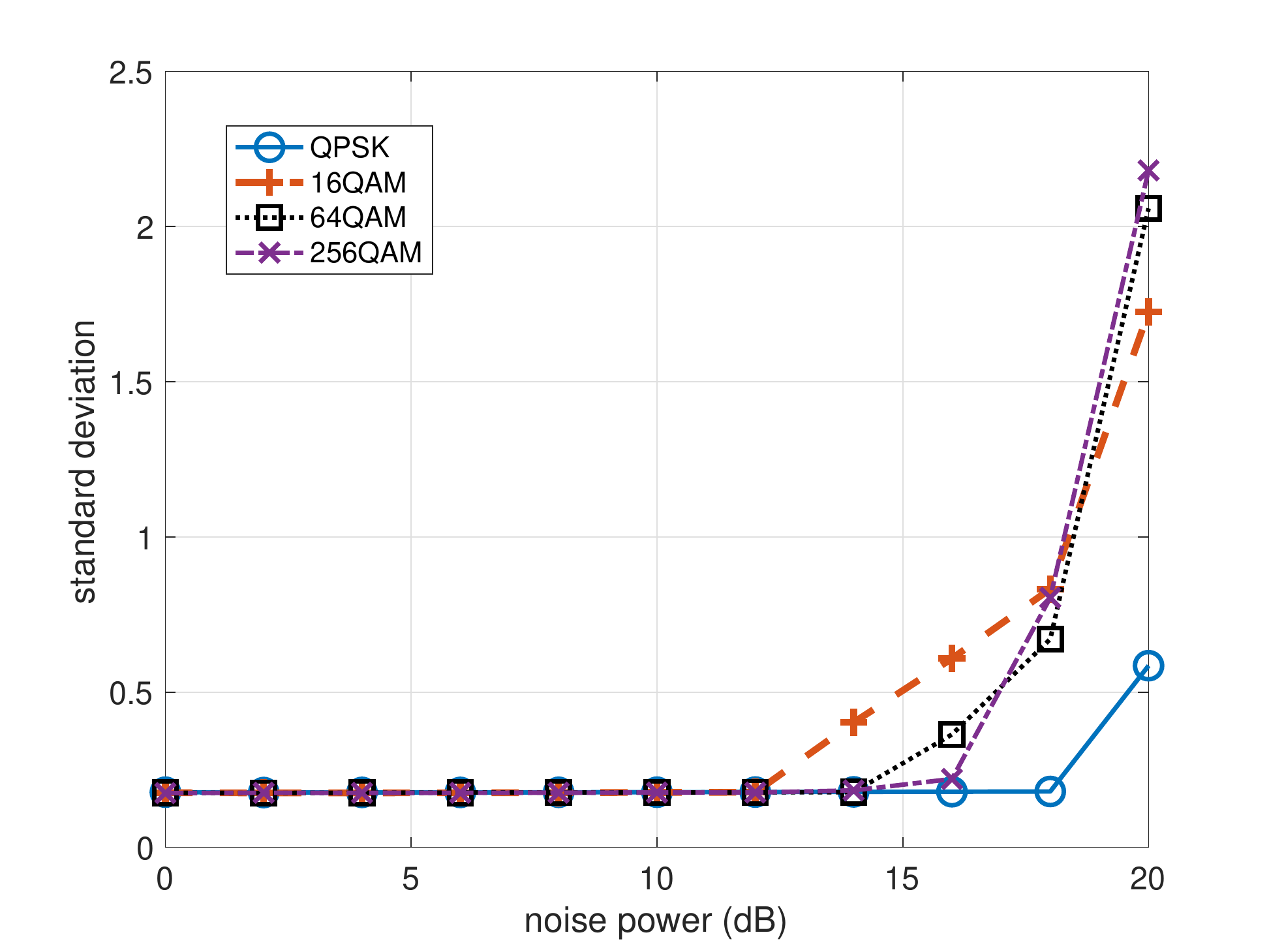}
  \caption{Ranging MSE with different noise powers and modulation schemes}\label{fig:mod}
  \vspace{-0.1in}
\end{figure}

The ranging performance with respect to different sampling frequencies is shown in Fig. \ref{fig:samp}. We observe that, when the noise power is significant, the higher sampling rate will substantially improve the performance, as being expected. The impact of different modulation schemes is shown in Fig. \ref{fig:mod}. We observe that the higher modulation order is, the worse the ranging performance is, when the noise power is significant. The QPSK, which has a constant envelop, achieve the optimal performance. It implies that the variation in the signal amplitude cause the performance degradation of ranging. 

\subsubsection{Communication-Sensing Tradeoff}

\begin{figure}
  \centering
  \includegraphics[scale=0.4]{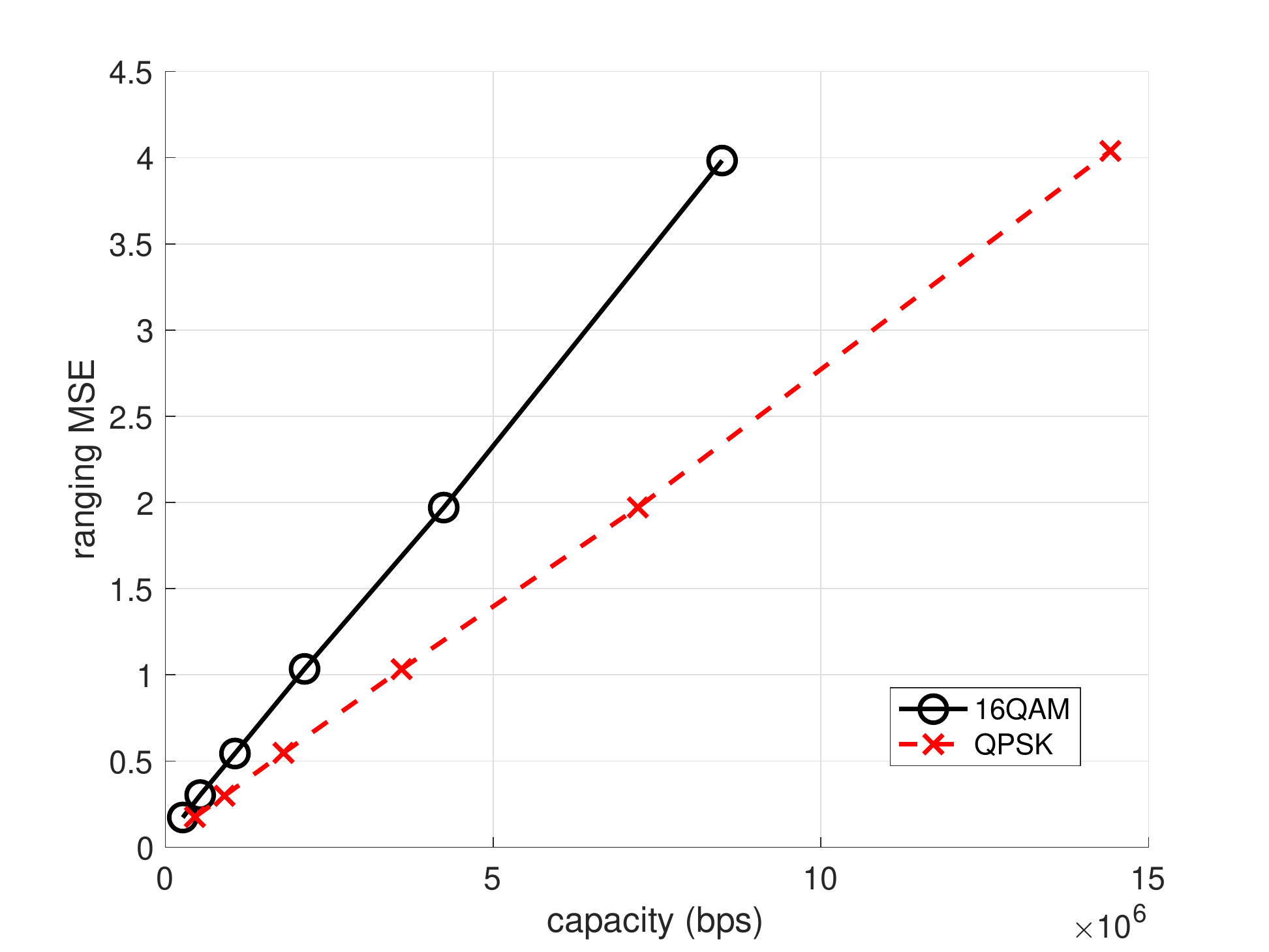}
  \caption{}\label{fig:tradeoff}
  \vspace{-0.1in}
\end{figure}

From the above simulation results, we observe that the order of modulation, which determines the communication data rate, has marginal impact on the sensing performance, unless the noise power is very high. Hence, the major tradeoff between communications and sensing is due to the implicit sharing of bandwidth as shown in Section \ref{subsec:tradeoff}. In Fig. \ref{fig:tradeoff}, we show the tradeoff between communication and sensing when different bandwidths are allocated to the communications while the total bandwidth is kept a constant. The modulation schemes of 16QAM and QPSK are tested. We observe that an increasing channel capacity implies worse performance of sensing (higher ranging MSE). An interesting observation is that the tradeoff curve is better for the QPSK modulation scheme. One possible reason is that the QPSK scheme has a constant envelop, while the phase change does not affect the performance of sensing. When QAM is used, the variance of signal power (thus the SNR) has a negative impact on the ranging performance. Essentially, this is due to the Jensen's inequality: as shown in Prop. \ref{prop:MSE}, the Cramer-Rao bound $MSE_{cr}$
\begin{eqnarray}
MSE_{cr}=E\left[\frac{C_0}{\gamma}\right]\geq \frac{C_0}{E[\gamma]},
\end{eqnarray}
where $C_0$ is the constant coefficient and $\gamma$ is random due to the different power at different symbol periods.

\subsection{FSK-SF}

\begin{table}[]
	\centering
		\caption{Configuration of simulations for FSK-SF}
		\label{table:para2}
	\begin{tabular}{|l|l|l|l|}
		\hline
		distance  & 100m & chirp duration $T_p$ &  60us\\
		\hline
		symbol number $N_s$ & 8 & modulation   & 8-FSK \\
		\hline
		freq. levels  & 512 & samp. freq.  & 136Msps   \\
		\hline
	\end{tabular}
\end{table}

\begin{figure}
  \centering
  \includegraphics[scale=0.4]{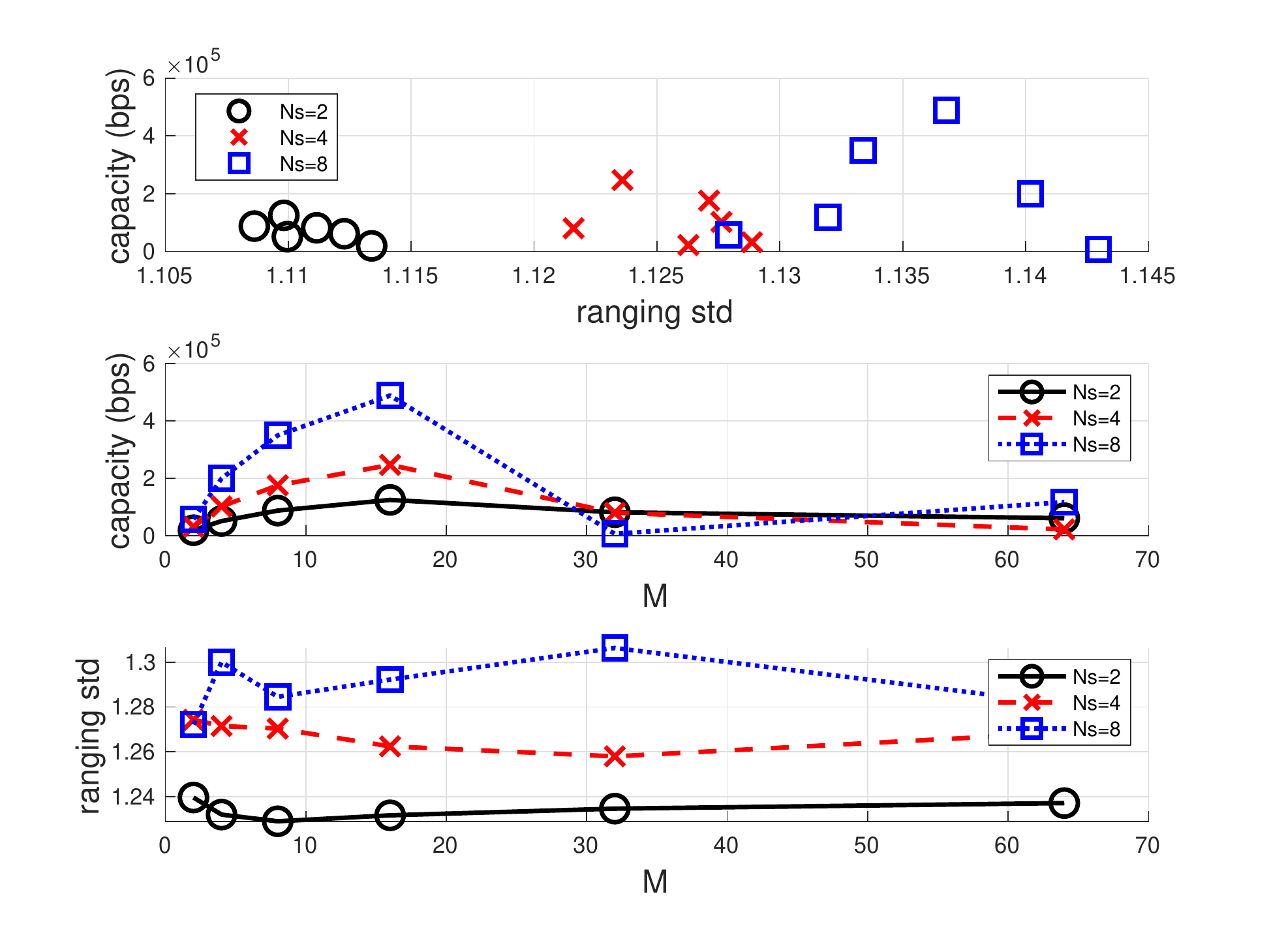}
  \caption{Performances of FSK-SF with respect to different $N_s$ and $M$}\label{fig:fsk-sf}
  \vspace{-0.1in}
\end{figure}

We also tested the performance of FSK-SF, whose parameters are shown in Table \ref{tab:para2}. In Fig. \ref{fig:fsk-sf}, we have shown the performance of communication and sensing with respect to different values of $N_s$ and $M$. The upper figure shows the points characterizing the performances. In each cluster of points due to the same $N_s$, different points corresponds to different $M$'s. We observed that a higher $N_s$ incurs better communication performance but worse sensing performance. An optimal $M$ can be selected for the optimal tradeoff between communication and sensing. In the middle subplot, we observe that the modulation order $M$ needs to be an intermediate number in order to maximize the communication performance. It is shown in the lower subplot that the modulation order $M$ causes marginal performance difference for the ranging. 

\begin{figure}
  \centering
  \includegraphics[scale=0.4]{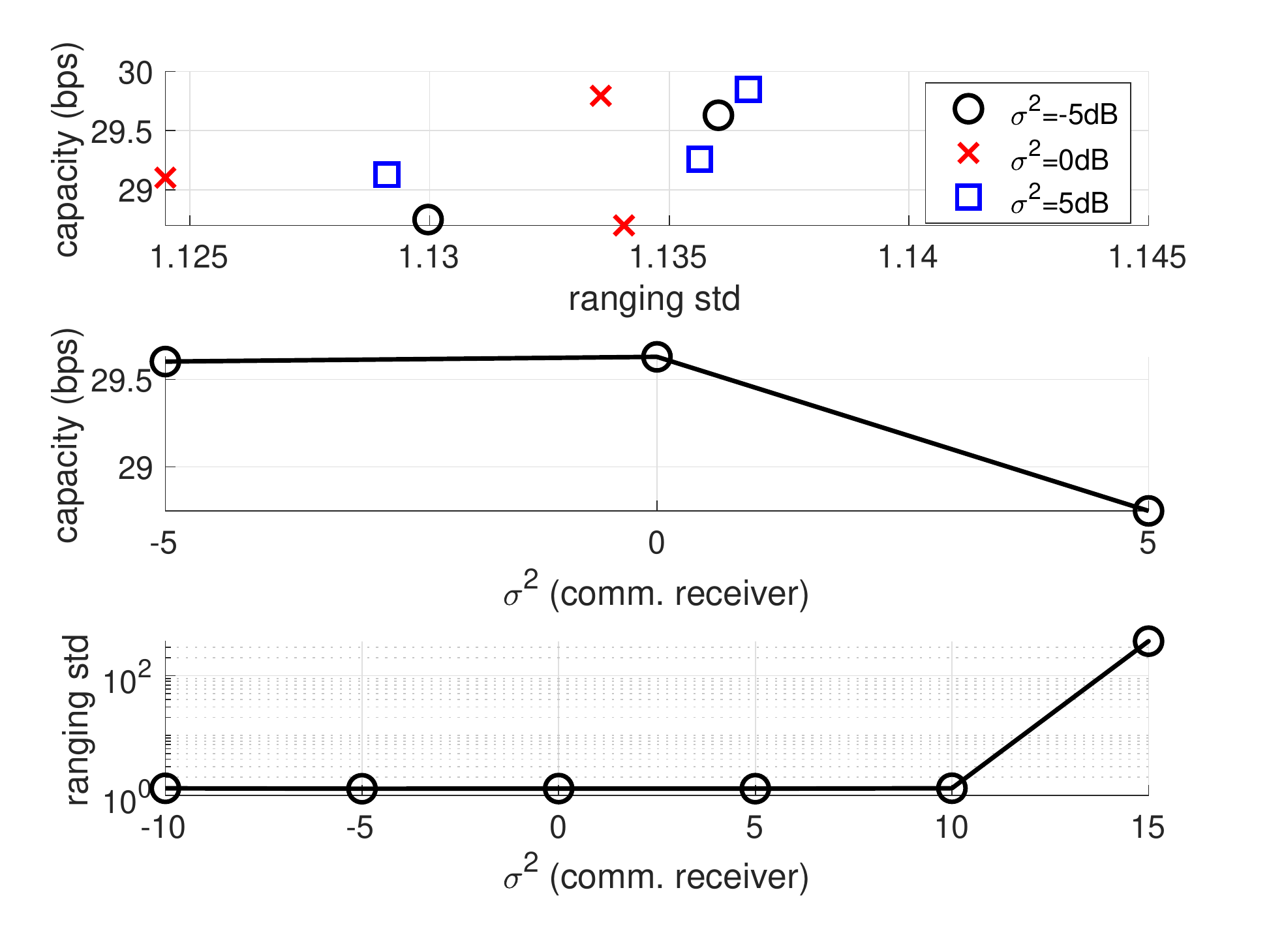}
  \caption{Performances of FSK-SF with respect to noise powers}\label{fig:fsk-sf_noise}
  \vspace{-0.1in}
\end{figure}

The performance of communications and sensing is shown in Fig. \ref{fig:fsk-sf_noise} for different noise powers at the communication and radar receivers. An observation is that the increase of noise power, either at the communication receiver or radar sensor, has marginal impact on the communications and sensing, unless the noise power is sufficiently high. 

\section{Conclusions}\label{sec:conclusion}
In this paper, we have studied the joint communications and sensing by leveraging the radar sensing waveforms. The approaches of QAM-FMCW and FSK-SF are proposed and analyzed. A major conclusion is that the total bandwidth used by the JCS is approximately the sum of the bandwidth used by communications and sensing, thus having no significant advantage over the scheme of separate communications and sensing, in terms of spectral efficiency.

\end{document}